# Functional surfaces through the creation of adhesion and charged patterns on azopolymer surface relief gratings.

Maria Gabriela Capeluto[a, *], Raquel Fernández Salvador[c], Rebeca Falcione[a], Aranxa Eceiza[c], Silvia Goyanes[b], Silvia Adriana Ledesma[a]

[a] *Laboratorio de Procesado de Imágenes (LPI), Departamento de Física, Facultad de Ciencias Exactas y Naturales, Universidad de Buenos Aires and IFIBA, CONICET, Cuidad Universitaria, Buenos Aires 1428, Argentina*

[b] *Laboratorio de Polímeros y Materiales Compuestos (LPM&C), Departamento de Física, Facultad de Ciencias Exactas y Naturales, Universidad de Buenos Aires and IFIBA, CONICET, Cuidad Universitaria, Buenos Aires 1428, Argentina*

[c] *Grupo "Materiales + Tecnologías", Departamento de Ingeniería Química y del Medio Ambiente, Escuela de Ingeniería de Gipuzkoa, Universidad del País Vasco, Pza. Europa 1, 20018 Donostia-San Sebastián, España*

*maga@df.uba.ar

**Abstract**

We show that an azopolymer can be used to create a supramolecular architecture in a parallel process with patterned surface properties. By illuminating with an interference pattern, we created adhesion and charge patterns that reflect the molecular ordering. We studied the recording process in two limit situations. When birefringence dominates over mass transport, an adhesion pattern was recorded even in absence of a surface relief grating (SRG). When mass transport dominates, we measured a higher frequency adhesion and relief patterns on top of the SRG. We measured an increased negative charge in the regions where molecules are expected to be parallel aligned in trans conformational state.

## 1. Introduction

Design and control of structure and properties of surfaces and interfaces is one of the challenges for material scientist and engineers. In general, the surface characteristics are related to the nature of the chemical and physical structure of the molecules at the nanoscale. In this context, some artificial materials are designed to perform specific functions at molecular level in view of the potential application on nanodevices. The following are only some examples of the wide variety of materials that had been designed. Graphene had shown to be an outstanding substrate for non-covalent supramolecular self-assembly functionalization to either control its electronic properties or provide new functionalities [1]. Chromophores were attached in a controlled way to metal or semiconducting surfaces for the construction of photonic devices and artificial surface-based light-harvesting systems [2]. Hierarchical assembly of one-dimensional nanostructures into well-defined functional networks had shown to form electrically conducting networks, with individually addressable device function at each cross point [3]. Ordered functional porous polymeric surfaces, has been designed with different functionalities inside the pores by using homopolymers, hybrid materials or chemical modification, with potential to applications in biomedicine and optics [4].

Molecular motions have attracted much attention as powerful tools for the construction of active materials or active components in smart devices. The photoisomerization of azobenzene molecules is especially interesting because, since their motions are consequent to photon absorptions, they can be remotely controlled by light without physical contact. Azobenzene is composed by two phenyl rings linked by a N=N double bond. When the azobenzene absorbs light of a proper wavelength, undergoes many reversible conformational cis-trans transitions; a property that gives the azobenzene molecule the capability of photo switching. The π electrons in the phenyl rings can interact non-covalently with a cation, an anion or other π system (π stacking interaction), so many functional groups can be weakly bonded to both ends of the azobenzene. The term π - π stacking is used to refer to all geometries of stacked interaction [5]. It is well known that face-centered stacking for azobenzene molecules is electrostatically unstable, while they can still interact through the edge-to-face (T-shaped or edge-on) geometry or the parallel displaced (or parallel off-centered).

The nanoscopic moieties of the azobenzene molecules can be used to produce microscopic or macroscopic scale changes in the properties of materials, when the azobenzene is employed in a host-guest or a functionalized azobenzene-containing polymer (azopolymer). The main properties observed when the material is illuminated by polarized light of a particular wavelength are induced birefringence and mass transport. Optically induced birefringence is produced under polarized illumination, because the cis-trans photoisomerization stops when the transition dipolar moment of the molecule is perpendicular to the polarization. Mass transport is a phenomenon not yet well understood, where mass on the surface of the azopolymer migrates to regions that are not illuminated. There is still an ongoing debate about the origin of the mass transport, some approaches tend to explain the phenomena by the existence of a photoisomerization force [6] while others by anisotropic light induced stress [7] or the photofluidization [8], the three of them initiated when the azo compound absorbs light. This very interesting light-matter interaction phenomena makes possible to engineer the azopolymer surface by light containing an intensity, phase or polarization distribution, to induce birefringence and produce mass patterns [9]. Azopolymer motions were used to organize CNT and re shape them, providing a platform for non-covalent functionalization by π staking interaction, both in the azo compound and in the CNT [10].

Mass pattern formation in azopolymers was extensively studied for more than 20 years [11,12], mainly through the illumination of interference patterns and consequent generation of the surface relief grating (SRG). Optical techniques had given information about the temporal evolution of both the orientation of trans isomers and the characteristics of the SRG. In the first case by measuring the optical induced dichroism (birefringence) with uniform illumination and in the second by measuring the diffraction efficiency in the SRG. Chromophore orientation across the SRG was studied by micro Raman spectroscopy [13,14].

While optical measurements were vastly used, surface sensitive techniques were somehow less explored. Yadavalli et al. combined optical techniques with AFM measurements to relate topography with different illumination conditions in PAZO and other azopolymers [15]. Nanoindentation was used to study the photomechanical properties of poly (disperse red 1 methacrylate) during the SRG formation, showing the photo-softening of the azopolymer due to the formation and expansion of free volume [16]. A similar study performed in poly (disperse red 1 acrylate) showed no substantial differences of photo softening between dark and illuminated areas on the sample indicating that material motion relies upon localized molecular scale rearrangements, that induces molecular pressure to distort the polymer matrix locally [17]. Electric Force Microscopy (EFM) allowed to measure the periodic electric polarization in periodically poled azo-copolymers [18,19]. Adhesion maps were used to characterize azobenzene single crystals and azopolymers films [20-22]. In single crystals, cis domains nucleate and grow after UV irradiation, displaying greater adhesion forces values for the cis isomer than for the trans isomer, differences that are explained from the inherent chemical heterogeneity of these isomers [20]. Azo/PMMA films exhibited different behavior highlighting that this is not a well understood phenomenon. In this case, upon UV illumination, the increasing content of cis isomer produces a diminished adhesion. The cis isomer has significant larger dipole moment than the trans isomer, that leads to a changed dipole-dipole interaction of the tip and the sample surface that influences the adhesion properties [21]. Adhesion and stiffness maps measured in cross-printed SRG, showed an increase of the stiffness in the crest and of the adhesion in the throughs. Differences in adhesion between the crest and throughs could be originated in the orientational distribution of chromophores that can impose different attractive interactions onto the cantilever [22].

In this work, we show that the mixed process of induced birefringence and mass transport, produces adhesion patterns and electrical charge distributions across the bright and dark fringes of the SRG. The first effect is probably related to differences in the organization of the trans isomer. The second one, consists of a negative charge enhancement in the illuminated areas, possibly originated in the π stacking interaction. To support this hypothesis, we present UV/visible spectroscopy measurements and we compare with birefringence measurements. The formation of these patterns is characterized for different exposures times and intensities. Adhesion patterns are found even in the early stages of the grating formation, when surface relief is still neglectable. The results are not only interesting regarding the possible applications but also could contribute to the understanding of the intriguing phenomena of mass transport. Molecular organizations could be a path to the development of azopolymer functional surfaces.

## 2. Experimental

### 2.1. Materials

We fabricated thin films from a mixture of a commercial azopolymer (PAZO), a plasticizer (ethylene glycol) and a solvent (methanol). Reagent grade methanol and ethylene glycol from Biopack (Argentina) were used in our experiment to develop the materials. Poly[1-[4-(3-carboxy-4-hydroxyphenylazo) benzenesulfonamido]-1,2-ethanediyl, sodium salt] (PAZO) was acquired from Sigma-Aldrich.

### 2.2. Film preparation and characterization

Azopolymer compound was fabricated by dissolving 100 mg of PAZO in 1mL methanol and 0.3 mL ethylene glycol using an ultrasonic bath for 30 min. The resulting homogeneous solution was deposited as thin layers on a clean coverslip by spin coating at a speed of 3600 rpm. We dropped a 100 μL volume of the solution, and let it spin for 1 min, repeating the process 3 times. The methanol in the samples was evaporated in a furnace, using a stepped rising temperature protocol that allows to produce homogeneous thin films without bubbles. The temperature sequence used was 12 h at 50 °C, 2 h at 60 °C and 2 h at 80 °C. The resulting film thickness of about 0.9 μm was measured from the cryogenically fractured surface of the samples using an Scanning Electron Microscopy (SEM). We measure the UV–visible spectrum which consist of a wide band peak centered in 360 nm and FWHM of 100 nm [10]. In order to isomerize the azo compound, we used photons at 473 nm wavelength from a 50 mW laser.

Samples were characterized by Scanning Electron Microscopy (SEM), Atomic Force Microscopy (AFM) and Electrostatic Force Microscopy (EFM). For the SEM measurements, we sputtered a thin layer of platinum (10 nm thick) to avoid charging of the polymer film while electrons hit the sample. AFM and EFM measurements were perform following the procedure of J. Gutierrez et al. [23].

Atomic force microscopy (AFM) was used to study the topography of the SRG. Adhesion maps were also acquired to characterize molecular states on the surface of the sample. The adhesion map is the result of a force-distance experiment in each point of the sample. AFM was performed using a Bruker Dimension Icon AFM operated under Peak Force mode in ambient conditions, with an integrated Bruker TAP525A tip having a resonance frequency of around 525 kHz, a spring constant of 200 N/m and a radius of 8 nm. Scan rates ranged from 0.9 to 1.1 Hz·s-1. Measurements were performed with 512 scan lines and a target amplitude around 0.9 V, in areas of 10 μm × 10 μm. Topography and adhesion maps were obtained simultaneously. We scanned different regions of the samples, to ensure repeatability of the results.

Electrostatic force microscopy (EFM) was used to study the electric properties of the azopolymer film. EFM measurements were also carried out under ambient conditions using the same scanning probe microscopy operating in lift mode (lift height was 300 nm) and equipped with an integrated Bruker Co/Cr-coated MESP tip having a resonance frequency around 75 kHz, a spring constant of 2.8 N/m and a radius of 35 nm. The secondary imaging mode, derived from the tapping mode that measures the electric field gradient distribution above the sample surface, was detected by applying voltage (12 V) to the cantilever tip. Locally charged phase domains on the sample's surface were qualitatively mapped simultaneously with the height measured at 0 V.

### 2.3. Birefringence

Birefringence was induced by illuminating the sample by a linear polarized 473 nm laser, with an incidence angle normal to the sample, as schematized in Fig. 1. The beam was filtered and expanded using a spatial filter composed by a 10x objective Melles Griot (O), a lens (L) and a 50 μm diameter pinhole (PH). The FWHM of the writing beam measured with a knife edge test was 2 mm. The time dependence of the optically induced birefringence was measured by illuminating the sample at an angle smaller than 5° with a linearly polarized 632.8 nm He-Ne laser (FWHM ~0.5 mm). Both lasers were chosen considering the absorption spectra of the azopolymer. In the optical path of the He-Ne laser, the sample was placed between two cross polarizers. A half waveplate was used to increase the transmission of the 473 nm laser on the polarizer (Po). The transmission axis of the polarizer P1 was placed at 45° from the polarization axis of the 473 nm laser given by Po.

### 2.4. SRG recording

As we have reported in previous papers [10,24] a standard Lloyd's mirror interferometer was used to generate an interference fringe pattern over the sample. A sinusoidal interference pattern is produced by the superposition of two half of the beam: one half that impinges on the sample after being reflected in the mirror and the other that impinges straight on the sample. The fringe period d = λ/2sin(θ) is determined by the incidence angle of the beam on the mirror (θ) and the wavelength of the light λ.

The complete schematic of the experimental setup used in this work is shown in Fig. 2(a). A 473 nm laser was filtered and magnified as in the birefringence setup. A linear polarizer (P) was used to configure the polarization of the interference pattern. A half wave plate (HWP) was used to optimize the intensity on the sample. The Lloyd's mirror (M) was set at an angle θ with respect to the incident beam direction. The sample (F) was mounted perpendicular to the mirror.

Considering full intensity illumination, the average power density on the sample was 0.2 W/cm2 approximately.

When the transmission axis of the polarizer (P) is set in the x direction (the laser is p-polarized) and the incidence angle is smaller than π/4, a large modulation of the intensity is obtained, keeping the polarization linear across the fringes. In this case, the two beams that interfere (one that impinges directly on the sample and the other that impinges on the sample after reflecting on the mirror) are p-polarized. The intensity distribution ($I^p$) for two p-polarized beams can be computed by using:

$$I^p = I_i(1 + \rho^2)[1 + \frac{2\rho}{1+\rho^2} \cos(2\theta) \cos(2k_o \sin(\theta) u)], \qquad (1)$$

where $I_i$ is the incident intensity, $\rho$ is the reflectivity of the mirror for photons of wavelength $\lambda$ at the angle of incidence $\theta$, $k_o = 2\pi/\lambda$, and $u$ the coordinate along the sample. A schematic of the distribution of intensity in one period of the interference pattern can be observed in Figure 2 (b) as a blue/white horizontal bar, where the scale bar on right indicates the intensity scale varying from 0 (blue) to $I_{max}$ (white). It is also shown with black arrows on top of the intensity distribution, the polarization vector of the electromagnetic field on the sample. This kind of intensity and polarization distribution is expected to induce an SRG shifted 90º with respect to the intensity distribution, as it is schematized in the yellow/black bar in Figure 2(b). The scale bar on the right of the yellow/black bar indicates the height scale varying from 0 (black) to $h_{max}$ (yellow). The net mass flow direction expected to produce this SRG is indicated in black arrows on top of the SRG schematic. We also indicate the points where there is not net mass flow (fixed points). As it will be explained later, since the velocity field is proportional to the intensity first derivative, the fixed points are located at the maxima and the minima of the intensity.

A 632nm laser was used to monitor the SRG formation by measuring the intensity of the diffraction orders 0 and 1 ($I_o$ and $I_1$ respectively). The diffraction efficiency was calculated as the ratio of the diffraction order intensities: $I_1/I_o$.

## 3. Results and discussion

Azopolymer SRG was imprinted on a large area of the film of about 14 mm2. We illuminated the sample with an interference pattern produced by placing the Lloyd's mirror at an angle of 5° from the direction of the incidence beam obtaining a fringe period of about 2.7 μm. The spatial light intensity distribution used for illumination consists of an interference pattern with the lines parallel to the flat edge of the mirror.

Fig. 3 shows SEM images of a typical SRG that was recorded during 14 h. As it was expected, the period of the SRG measured in a top view of the sample (Fig. 3(a)) is about the period of the interference fringes. A fracture of the material in each height minima occurs in some of the samples near the flat edge of the mirror where the intensity is higher (black line in Fig. 3(a)). This fracture might be produced by the induced stress during the mass transport. Notice that the fracture occurs in the height minima of the SRG, where there is a node for the net mass flow (Fig. 2(b)), from where the mass if flowing outward pulling the material towards the SRG maxima. The nonlinear response of the material is revealed from the SEM image of a cryogenic cut of a sample (Fig. 3(b)) [10], in a region further away from the edge of the mirror, where the fracture is not observed but the grating has a triangular profile. A sinusoidal profile would be expected if the response were linear.

In addition to the SRG, a smaller subperiod pattern is observed. The measurement using the e-beam impinging normal to the substrate makes evident the smaller undulations that are seen as different grayscale lines on top of the SRG in Fig. 3(a). To make these undulations more evident, we show an averaged profile (inset in Fig 3(a)) for 4 undulations from a small region of the SEM image (white line on Fig. 3(a)). The maxima positions of the secondary pattern are indicated with white arrows in Fig. 3(b). The modulation depth in this sample is about 990 nm (~2 λ) for the principal pattern and 15 nm for the much smaller secondary pattern [10]. As it will be explained in more detail later, this secondary effect was previously observed [25-28] and in particular we used it in a previous work to change the shape of a CNT [10].

During the time the grating was recorded (14 h), diffraction efficiency grew continuously without saturation (Figure 4 (a)). For the same writing beam intensity, the optically induced birefringence (Δn) reached it saturation only after 50 minutes of exposure to light (Figure 4 (b)). Therefore, at equal illumination conditions, birefringence saturates much faster than mass transport. As it is well known, not only the intensity is important to activate mass transport but also the intensity gradients and radiation pressure. Thereby, in the limit of very weak intensities or very short exposure times, birefringence is expected to predominate over mass transport (times shorter than 50 min).

A lower intensity beam will produce a smaller rate of photo isomerization transitions, and consequently it will take more time to reach the saturation of the photoisomerization. Figure 4 (c) and (d) show birefringence and the UV/Visible absorption spectra during the process of birefringence saturation, thermal decay or relaxation and erasing using for the recording process a laser intensity of 18 mW. To avoid saturation of the absorbance, in this experiment we used a sample fabricated by spin coating only one layer of the azopolymer. Before illumination, the sample is isotropic (random oriented trans isomer). The UV/vis absorption spectrum is composed by strong absorption band due to the π-π* transition in the trans isomer (black curve on Figure 4 (d)). The peak absorbance is about 3.5 at 358 nm. The initial concentration of trans isomers in dark conditions [trans]$_{dark}$, is the maximum number of molecules per unit of volume available to isomerize. From Beer Lambert law, [trans]$_{dark}$ is related to the peak of the absorbance in the UV/vis spectra $A_{dark}$ as $\epsilon_{dark} d$[trans]$_{dark}$ with $\epsilon_{dark}$ the molar concentration and $d$ the thickness of the sample. Saturation of the birefringence, using a thinner sample and a lower power laser is obtained after 30 min. Then the maximum possible trans isomers are aligned perpendicular to the polarization of the laser. The total concentration of trans isomers (oriented or not) [trans] and cis isomers [cis] should be equal to the initial concentration of trans isomers in dark conditions [trans]$_{dark}$. After turning off the illumination, birefringence reaches its stationary value with a characteristic time $\tau_{th}\sim$ 1h so a proportion of the oriented trans isomers are now randomly oriented due to relaxation, this proportion is related with the difference between the maximum and the remnant birefringence. We measure the absorption spectra during relaxation in dark conditions. We performed measurements right after the laser was turned off, every 5 minutes during 2 h. Since $\tau_{th}$ is smaller than the experiment time length (2 h), an increase in the absorption due to the increase of trans isomers randomly oriented is expected. However, even observing a decrease in the birefringence, we obtained the same spectra for each time (red curve in figure 4 d)): a decreased absorption peak related to the photo orientation of trans azobenzene and cis remaining isomers. It is maybe possible that birefringence decreases because material thermal relaxation, but the trans isomers concentration doesn't change because there is a significant amount of trans isomers that had π stack after irradiation. Conventional azobenzene are planar molecules that can easily π-π stack, in particular it had been shown a strong effect in reversible memory storage using PAZO [29, 30].

The erasing process consisted of illuminating the sample with circular polarized (CP) light at 473 nm. As it can be seen in Fig. 4 (c), full decay was reached during the illumination time, meaning that trans isomers are randomly oriented. The absorption measured immediately after CP light irradiation increases but it doesn't fully go back to the initial state (trans with no orientation). There are many factors that may contribute to the increase in the absorption peak after erasing, other than the cis isomer thermal back conversion. On the one hand, π staking is an attractive weak interaction that competes with the dipole Coulombic repulsive interaction between the parallel dipole moments of the trans isomers [11]. On the other hand, the energy associated to the interaction is rather low [31,32], so photons in the blue spectral range can easily break the dimer formed by the two azo compounds. In any case, there is an increased amount of trans random oriented isomers.

A detailed inspection of the sample using different techniques could reveal the dynamics of the molecular organization on the azopolymer SRG. We analyze the measurements of topography, adhesion and electrical charge distribution for different conditions of illumination, by changing the intensity and exposure time.

First, we study the characteristics of samples fabricated under low intensity conditions. We illuminate the sample with an attenuated beam using a neutral filter (OD 3.3). Fig. 5 show topography and adhesion maps, for samples that were illuminated by the weak interference pattern for 10 min and 14 h respectively. As it can be clearly seen from the topography images for both recording times (Fig. 5 (a, c)), there is not a SRG recorded on the surface. This could possibly happen not only because of the low intensity, but also because of the shallow intensity gradients. However, adhesion measurements exhibit periodic patterns with 50 nN and 100 nN modulation depth for the samples that were exposed by 10 min (Fig. 5 (b)) and 14 h (Fig. 5 (d)) respectively. The measurements in Fig. 5 (b) were filtered using a FFT band pass filter to remove noise. Even for the shorter recording time (10 min), where mass transport is expected to be negligible, there is evidence of a surface adhesion pattern.

These periodic adhesion distributions can be attributed to the existence of spatial regions with molecules in different states, constituting the birefringence pattern, as it is schematically shown in Figure 5 (e). This can be understood from the fact that, as it was explained before, birefringence saturates much faster than mass transport. Even though the effect of mass transport is not observed, differences in the direction of the molecular dipole transition moment ($\bar{p}$) between the dark and bright areas of the illumination are still expected. As it is well known, the more probable state of the azobenzene compound in dark conditions, regarding its structural and thermal stability, is the trans conformational state. Due to the anisotropic characteristic of the system, there will not exist a preferential orientation for $\bar{p}$. We call that situation as trans state with random orientation of $\bar{p}$ ($\bar{p}_{rand}$). When the azopolymer is illuminated by an interference pattern with linearly polarized photons whose wavelength falls in the absorption band of the absorption spectra, there will be mainly two different regions. The dark areas will remain trans with $\bar{p}_{rand}$ orientation. In the bright areas of illumination, the azo compound undergoes many cis-trans photo isomerization transitions, until $\bar{p}$ is oriented perpendicular to the electric field ($\bar{E}$) when the transition dipole moment is $\bar{p}_\perp$. Therefore, the illuminated areas are composed by molecules in *cis* and *trans* states, with an increasing population of molecules in trans state with $\bar{p}_\perp$ as the recording time increases. Saturation of the material birefringence means that the maximum possible number of molecules reach its final state (trans, $\bar{p}_\perp$), and the remaining molecules will be in a stationary situation where the rate of transitions from cis to trans equals the rate of transitions from trans to cis. Even when saturation conditions are not met, after turning off the illumination, this kinking process will stop and the molecules in cis state will thermally decay to the trans state. The most probable state will be trans, but there will be a remaining spatial pattern in phase with the illumination pattern with differences in the orientation of $\bar{p}$. In summary, previously illuminated areas will be trans with preferential $\bar{p}_\perp$ orientation, while dark areas will be trans with $\bar{p}_{rand}$ orientation.

There are many factors that had been suggested to impose different attractive interactions onto the cantilever. The orientational distribution of chromophores [22] and the inherent differences in the chemical nature of trans and cis isomers [20] had already shown differences in adhesion. The local macroscopic Polarization of the sample considered as the sum of the electric dipole moments of the trans isomers per unit of volume changes across the SRG: it is minimum where they are randomly oriented and maximum where they are parallel aligned. Differences in Polarization had already shown sensitivity in adhesion in periodically polled azopolymers [18,19]. Also, since the azobenzene molecule is covalently bonded to the polymer chain, and its trans isomer is a planar molecule that can easily π- π stack [29], this preferential orientation will lead to a more compact packing of the azobenzene molecules (compared to non-interacting molecules) in the illuminated areas. Strong π-π stacking of azobenzene chromophores may lead to the reduction of free volume for the azobenzene to isomerize [30]. We would like to stress out that this is a transitory situation before mass transport is significant. The possibility of generating these adhesion patterns is very interesting because they can be used, for example, to create molecular patterns or set a platform for molecular decoration [33].

As a second study, we characterized the samples fabricated under a higher intensity (laser without attenuation), which allow us to induce a SRG. Fig. 6 shows the topography and adhesion maps of samples that were illuminated for 10 min (Fig. 6 (a) and (b)) and 14 h (Fig. 6 (c) and (d)). The topography images show recorded SRGs that are in good agreement with the SEM image from Fig. 3. When illuminating for 10 min, a 100 nm modulation is measured, and a double-peak structure is observed in the crests of the SRG. For longer time (Fig. 6 (c)), a modulation depth of about 950 nm is measured from the AFM image. The smaller sub-period pattern observed in the SEM images is not easily seen in the AFM image because the main pattern and sub pattern have heights in very different scales. However, adhesion maps show a patterned surface with smaller period than the SRG grating, that are surprisingly noticeable in Fig. 6(d). In this case, mass transport and molecular orientation are competing processes: in the areas with higher illumination, molecules are predominantly oriented with $\bar{p}_\perp$, however the mass is dragged from the bright areas to the dark areas and it is accumulated at the crests of the SRG. We have previously shown that adhesion can be used to distinguish regions where azobenzene has predominant orientation of transition dipole moment and higher azopolymer packing. This measurement could indicate that the small secondary pattern is not only a simple small relief, but it is also composed by regions with different azopolymer molecular states.

The secondary patterns of sub-wavelength characteristic sizes were observed before in SEM images, however in most of the cases they were not discussed [25] or even not commented at all [27,28]. Pawlik et al. observed the build-up of a double-peak structure and explain it phenomenologically from the cancellation of two mass currents that runs in opposite directions (one due to the gradient of illumination and the other due to Fick's first law of mass transport, and consequently in the density of mass gradients) [26]. As it is represented in Fig. 2(b), it should be also notice that, since velocity is proportional to the first derivative of the intensity [26], there are locations where mass is not transported (fixed points for mass transport): the locations at the intensity maxima, and the locations at the intensity minima. Following this reasoning we could explain the small subwavelength pattern as a standing wave between the fixed points originated in the two-mass current flowing in opposite direction. From another point of view, Sekkat proposed the existence of a photo isomerization force, with different characteristics of the gradient force. Both forces act on the polymer and compete with elastic forces and random forces due to spontaneous diffusion [6]. Gradient force can change direction depending on the ratio of the refraction index of the material to the surrounding media (air in our case). This force acts only in the polymer surface since it requires an interface between two media with different refraction index. Sekkat states that when photoisomerization and gradient forces have same directions, they could interfere in constructive or destructive way. This can also support our measurements, because if this interference occurs in a smaller scale than the SRG period, a periodic subpattern will build up. Despite the efforts made to explain the mass transport phenomena, there is still several arguing about the origin of the material deformation. This difficulty to explain the behavior of the material is probably from the fact that mass transport phenomena close to the surface is nonlinear [25,26] and, the secondary patterns can appear under certain experimental conditions, that might depend on polymer viscosity and elasticity [26]. Saphiannikova et al. [7] attribute the deformation to anisotropic light-induced stress due to the orientation of the azobenzene dipole moment perpendicular to the polarization direction. As a conclusion, either competing forces, opposite mass currents or internal stresses could contribute to the build-up of these secondary structures.

Because of the photoisomerization process and azobenzene interactions, a particular charge distribution can be expected. In fact, when birefringence is saturated and the transition dipole moment of most of the trans isomer are oriented perpendicular to the electric field (and then parallel between each other), the phenyl rings of different molecules have an increased probability to interact through parallel off-centered stacking over the edge-to-face geometry [5]. Therefore, charge differences should be expected between the bright and dark areas of the illuminated sample. To unveil this charge pattern, we used EFM measurements. Contrary to topography and adhesion maps, where the AFM tip gets in close contact with the surface, electrical properties of the sample are measured in lift mode, in such a way that there is not contribution from the van der Waals forces and other short-range forces. We characterize the electrical properties of the sample for the two limit conditions (low and high intensity) for samples illuminated during 14 h, as it is shown in Fig. 7. Fig. 7 (a) and 7 (b) correspond to the EFM images taken at 0 V and 12 V respectively, in an area that was illuminated with low intensity. As it can be clearly seen, no signatures of relief neither electrical charge are observed. However, it was previously observed in Fig. 5 (b) that a spatial distribution with different molecular states should appear. The fact that we have not measured charge in these samples, is probably because it is too low. Fig. 7 (c) and 7 (d) correspond to the EFM images taken at 0 V and 12 V respectively, for a sample illuminated with higher intensity. The measurements at 0 V are influenced by the relief of the sample, and the SRG maxima are visible. Electrical charge spatial distribution measured at 12 V is observed in Fig. 7 (d). As it was explained by Gutierrez, positively charged domains appear dark in the EFM phase image and the negatively charged domains appear bright. On one hand, by comparison of Fig. 7 (c) and 7 (d), it is observed that the SRG is about 90° shifted with respect to the electrical charge distribution. On the other hand, as it is well known, the SRG is shifted with respect to the birefringence grating. Then, the birefringence grating, the electrical charge grating, and the illumination are most likely to be in phase. Additionally, as it was explained before, molecules in the regions that were illuminated by the bright areas of the interference pattern, are mainly in trans state with $\bar{p}_\perp$. In principle, the planar π-system backbone of the trans isomer could be oriented in any angle around $\bar{p}_\perp$. However, since large π stacking interaction between phenyl rings from different molecules it is expected, they will interact with each other through parallel displaced (parallel off-centered) stacking [5]. Therefore, π stacking interaction between phenyl rings produce large charge delocalization that is measured macroscopically as an increased negative charge of the sample. This electrical charge distribution provides a well-ordered patterned surface for functional groups that could be able to be engaged through electrostatic interactions.

Fig. 8 shows the general picture of adhesion and charge pattern formation. The blue background is a section of the interference pattern used for illumination (in light blue the highest intensity). The polarization vector across the interference pattern is indicated in the lower part of the blue background with black arrows. The azo compound orients with the transition dipole moment normal to the polarization vector in the brighter areas of the interference pattern, while it remains random oriented in the darker areas. The azo compounds have higher probability to π stack in the bright areas where their planar structures are aligned (azo compounds in blue dashed ellipses). The adhesion and charge pattern are represented as green and yellow lines in the bottom respectively. Higher adhesion and more electronegative charge distribution, represented with lighter colors, are found in the bright fringes of the interference pattern, where molecules have higher degree of organization, the π stacking of trans azobenzene is more probable, and the electric polarization vector is bigger.

## Conclusions

We have demonstrated that the recoding of SRG involves the creation of adhesion and charge patterns originated in the dynamics of molecular transport and orientation. These patterns could have important applications being platforms for functionalization through electrostatic bonding. Adhesion was observed even in absence of a SRG grating. This result is in accordance with the fact that birefringence saturates faster than mass transport. Sub wavelength patterns were clearly observed in topography measurements. The sub-wavelength patterns build-up was discussed taking into consideration the dynamics of the mass transport. Either competing forces, opposite mass currents or internal stresses could contribute to the formation of these secondary structures. Regarding the EFM measurements, higher charged areas were found also in the areas of higher illumination, where the most probable state is $\bar{p}_\perp$ and large π stacking in parallel shifted configuration is expected. Considering that azopolymer films can be driven by light to produce 3D patterns [9], recording volumetric structures could be a new strategy towards 3D functional surfaces for nanotechnology.

## Acknowledgments

The authors would like to thank the financial support from the University of Buenos Aires (UBACYT 20020170100564BA and 20020170100381BA), ANPCyT (PICT 2017-2362, PICT-2014-3537, PICT 2014-2432) and the Basque Government (IT-776-13). The authors would also like to thank the technical support from "Macrobehaviour-Mesostructure-Nanotechnology" SGIker unit from the University of the Basque Country (Spain), and the Quantum Electronics Lab (UBA, Argentina) for lending the instrumental.

**Figures**

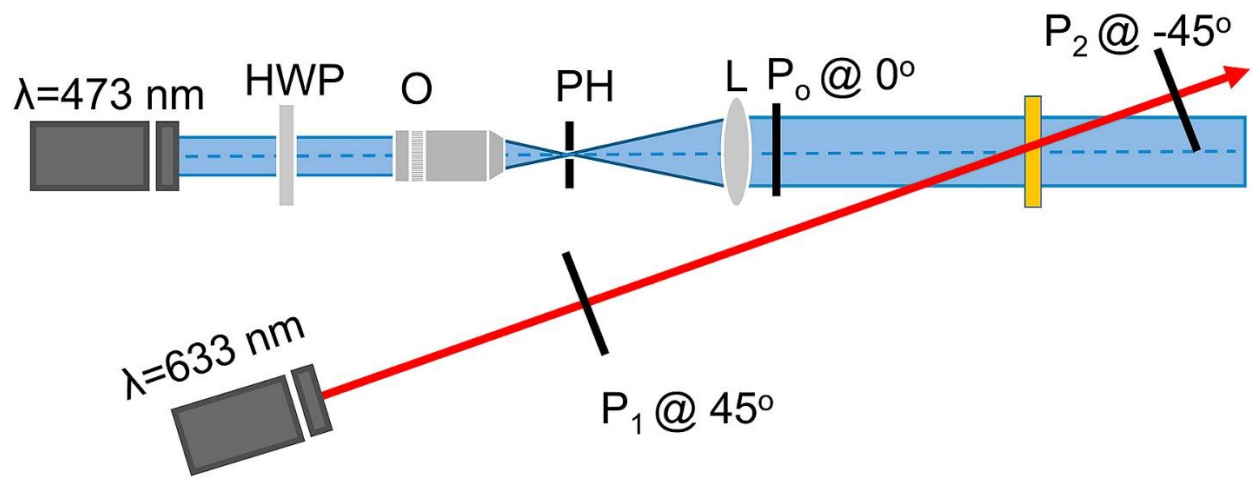

Fig. 1. Schematic of the setup used to measure the induced birefringence. HWP: Half Wave Plate; O objective; PH: pinhole; L: lens; Po, P1, P2: Polarizers; S: Sample.

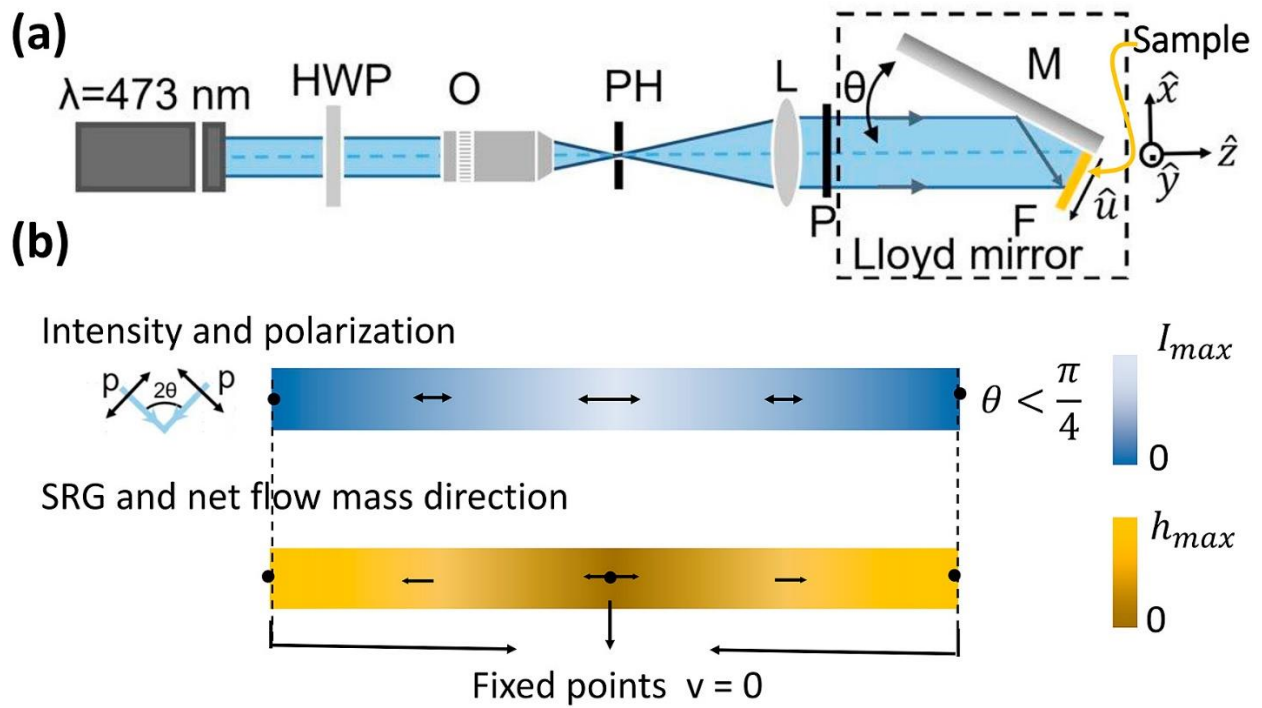

Fig. 2. (a) Experimental setup. HWP: half wave plate, O: microscope objective, PH: pinhole, L: lens, P: polarizer. M: mirror (b) Schematic of the polarization (black arrows) and intensity distribution (blue colormap bar) for a linearly p-polarized beam, in one period of the interference pattern, and the induced SRG (yellow colormap bar) indicating the net mass flow direction (black arrows) and the fixed points.

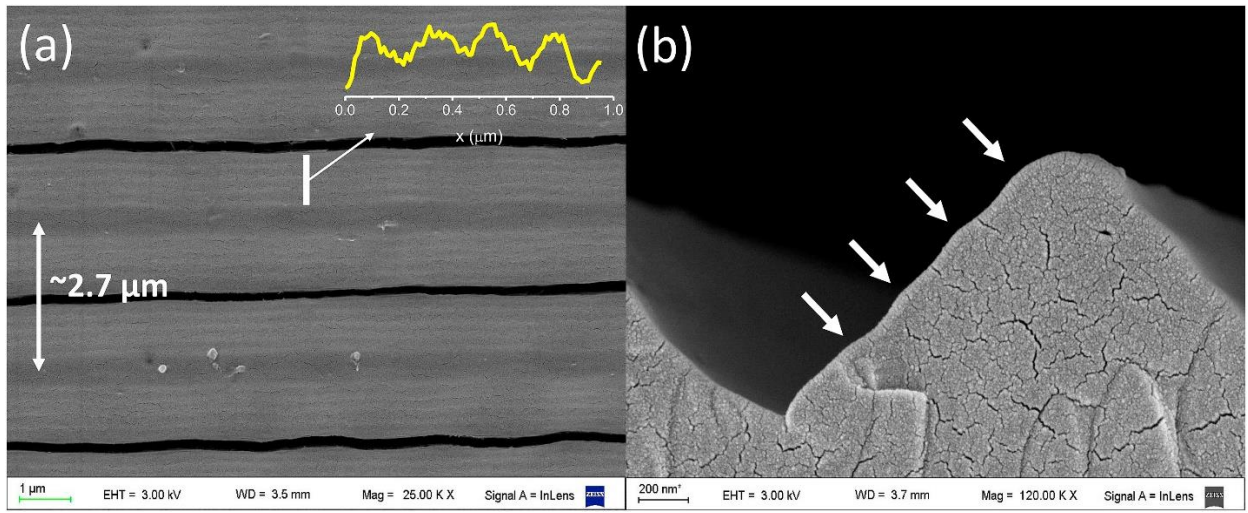

Fig. 3. SEM images of a typical sample. (a) SEM image measured from the top of the sample reveals a smaller period line pattern parallel to the main SRG lines, and a fracture in each height minima. The inset on the top right shows an average profile of the image that makes evident the smaller subperiod pattern. (b) A magnification of the transversal cut of the SRG shows the two-scale line pattern. The arrows indicate the subperiod pattern maxima.

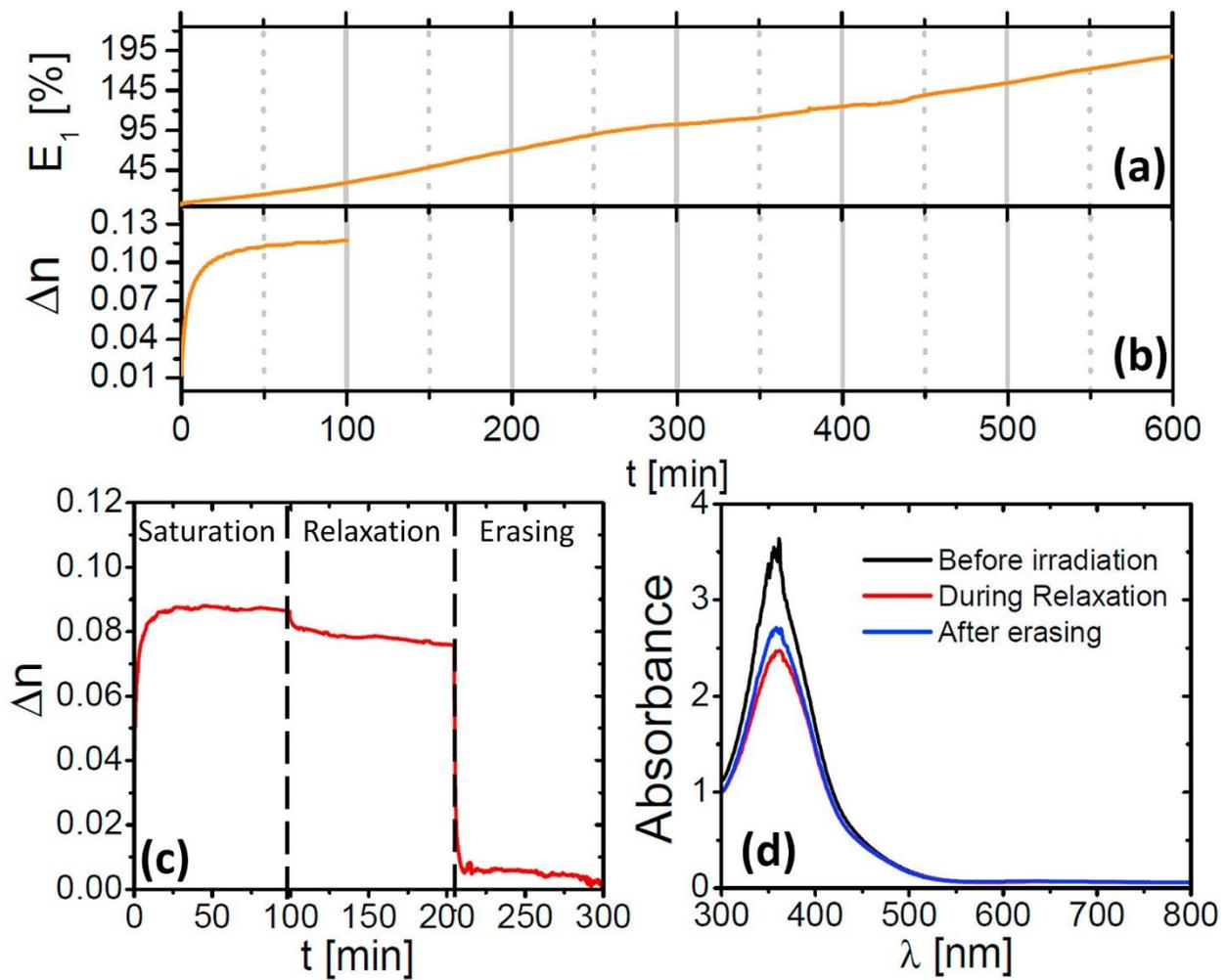

Fig. 4. Temporal evolution of the diffraction efficiency (a) and optical induced birefringence (b). (c) Saturation, relaxation and erasing of the birefringence, and the corresponding UV/vis absorption spectrum (d).

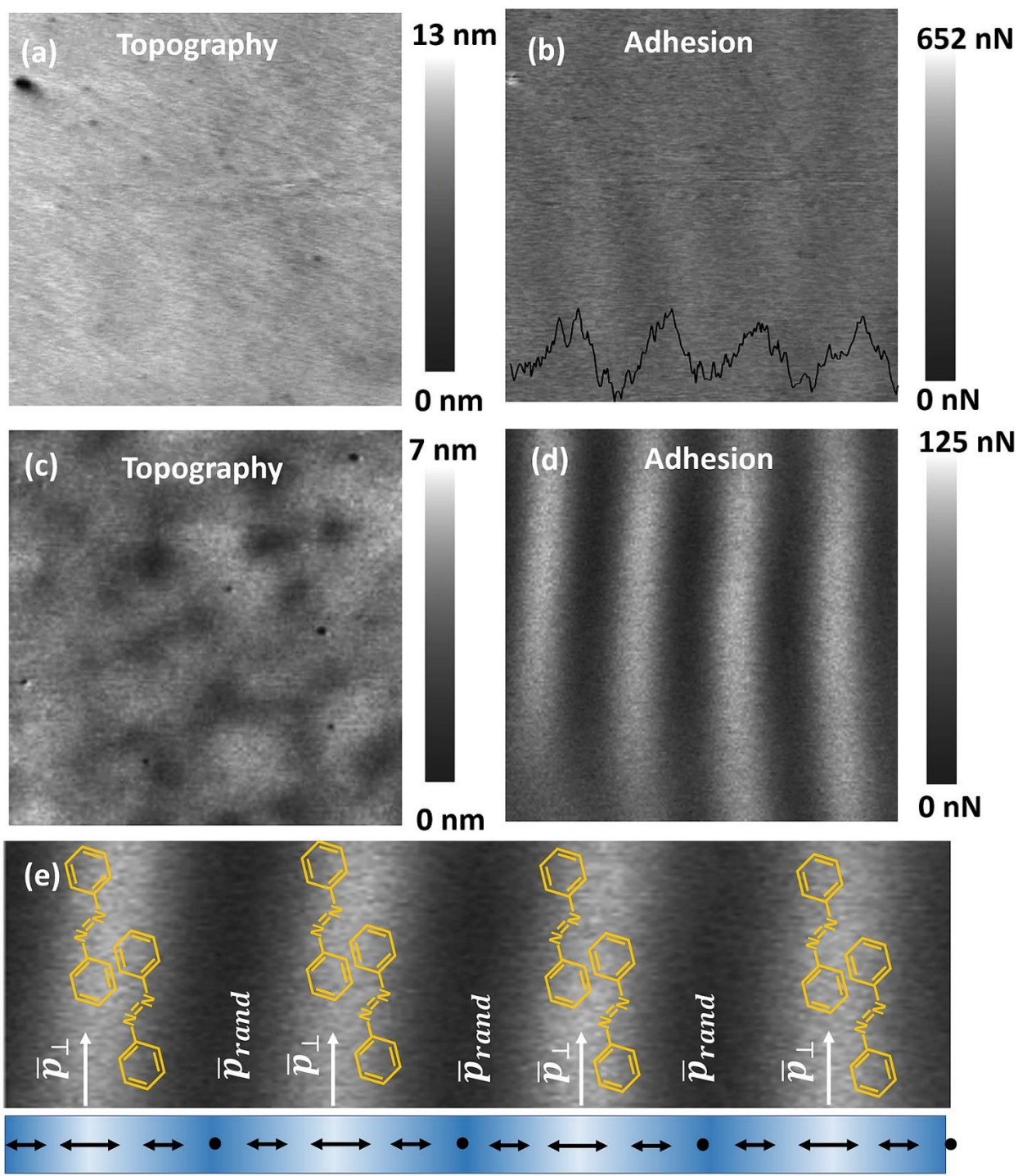

Fig. 5. AFM images of a sample illuminate 10 min and 14 h. (a) Topography (b) Adhesion measurements for the sample illuminated for 10 min and (c) Topography and Adhesion (d) for the sample illuminated 14 h. (e) Schematic of the most probable molecular states and transition dipolar moment across the sample, and its correspondence with the intensity pattern.

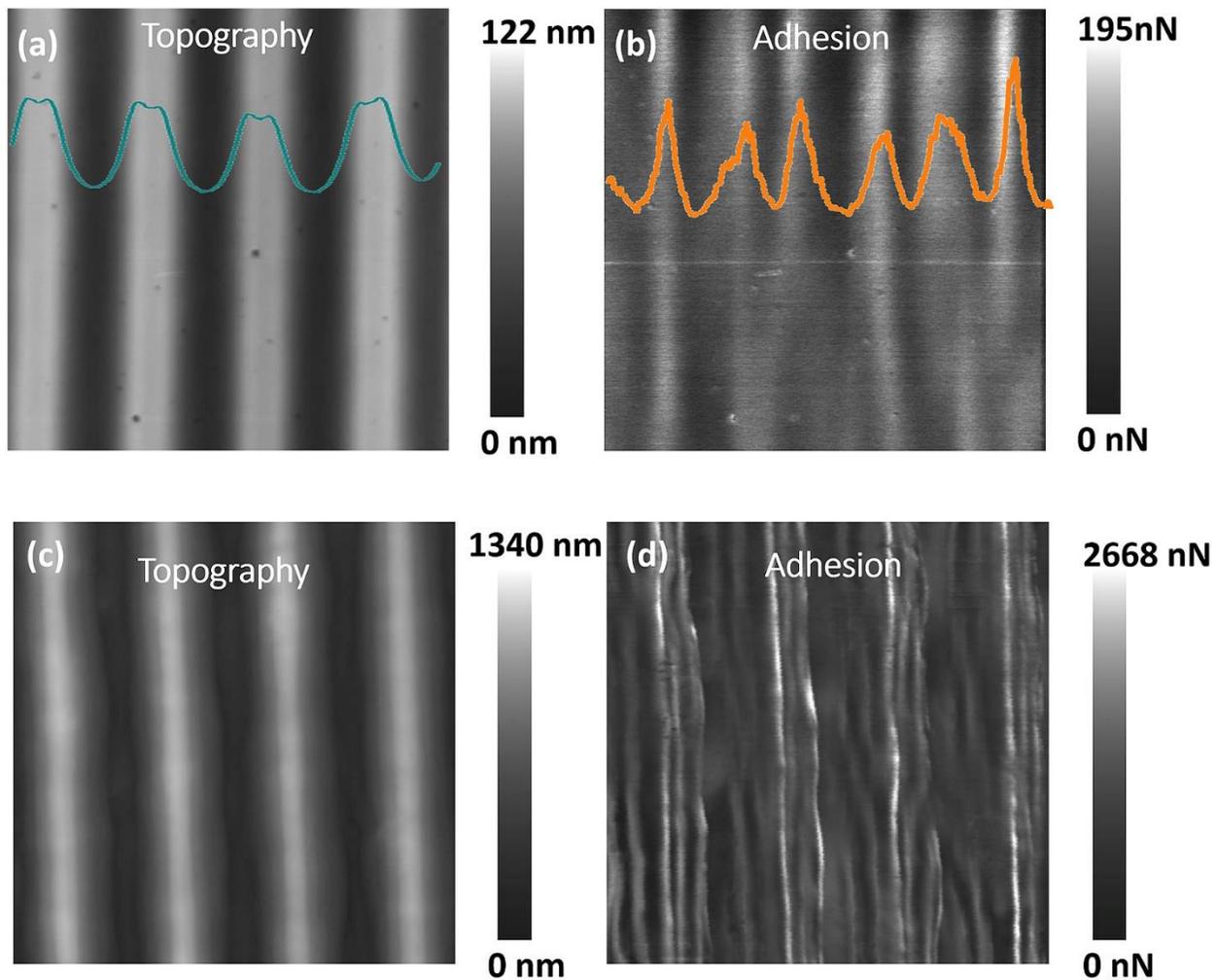

Fig. 6. AFM images showing topography and adhesion of a sample that was illuminated for 10 min (a, b) and a sample that was illuminated for 14 h (c, d).

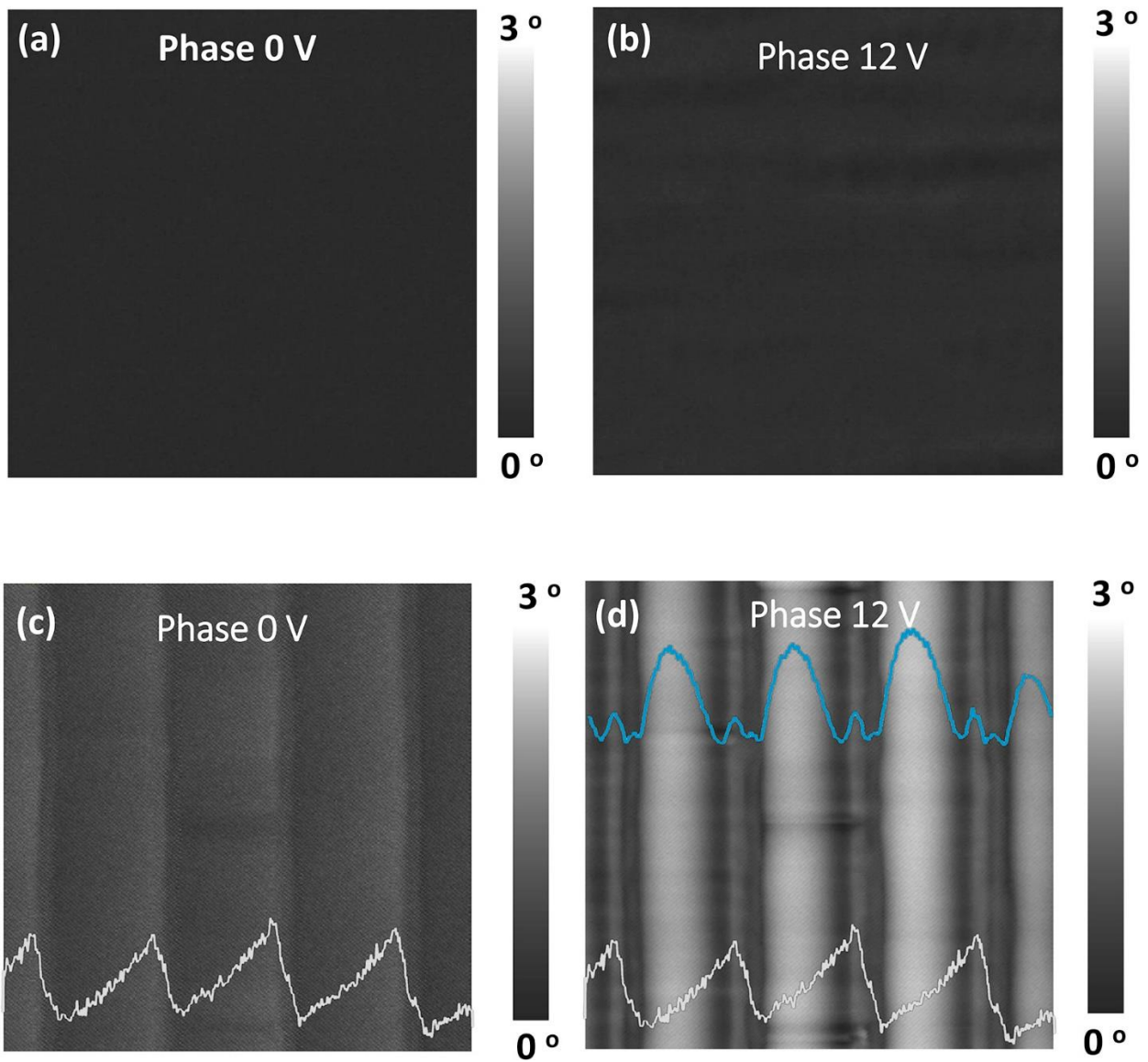

Fig. 7. EFM images at 0 V and 12 V for a sample illuminated with the attenuated laser (a and b)) and a sample illuminated with full intensity (c and d). In c) and d) we draw the height profile for 0 V (white line) to show that the charge profile (blue line) in (d) is shifted 90°.

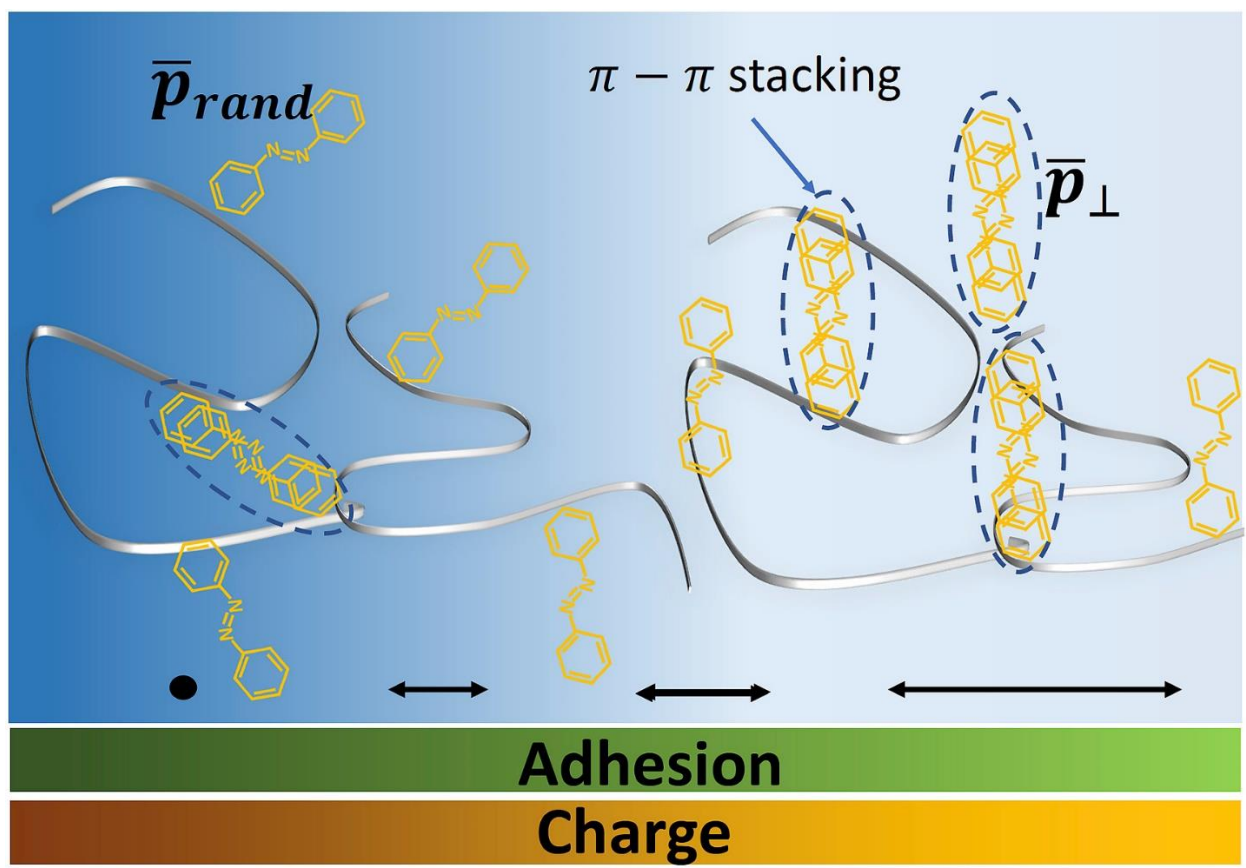

Fig. 8. Schematic of azobenzene transition dipolar moment orientation and azopolymer packing across the sample. The blue gradient represents intensity distribution (higher intensity is light blue), the polarization is represented with black arrows. Green and yellow bars represent adhesion and charge patterns. Light green means higher adhesion and light yellow means the highest negatively charged areas.

## Supplementary Material

Cis-trans kinetics characterization. Figure S1 (a) shows the UV/vis absorption spectra during irradiation with UV unpolarized light (365nm). It can be seen from the π-π* transition band (peaked at 358m) that there was not full conversion from trans to cis isomer during the illumination time. The peak absorption dependence on the irradiation time (red dots in Figure S(b)) determines the kinetics of trans to cis isomerization. The absorption maxima decreased due to the increased amount of cis isomer with a characteristic time roughly estimated on 500 min. To study the dynamics of the cis-trans back thermal conversion, we measure the evolution of the spectrum in dark conditions right after turning off UV illumination (black dots in Figure S(b)). The cis isomer fraction ([cis]) can be determined from the peak absorbance A at wavelength $\lambda$, by

$$[cis] = \frac{1 - A/A_{dark}}{1 - \epsilon_{cis}/\epsilon_{trans}} \qquad (1)$$

where $A_{dark}$ is the initial peak absorbance with only trans present at the wavelength $\lambda$, and $\epsilon_{cis}/\epsilon_{trans}$ the molar absorption coefficients of cis and trans isomers at $\lambda$, which is approximately 0.05 for azobenzene [1, 2]. Figure S2 shows the concentration of cis isomer during the trans→cis isomerization and during the cis→trans thermal back conversion. The remaining concentration of cis isomer after 20h of turning off illumination is on the order of 0.05 on average.

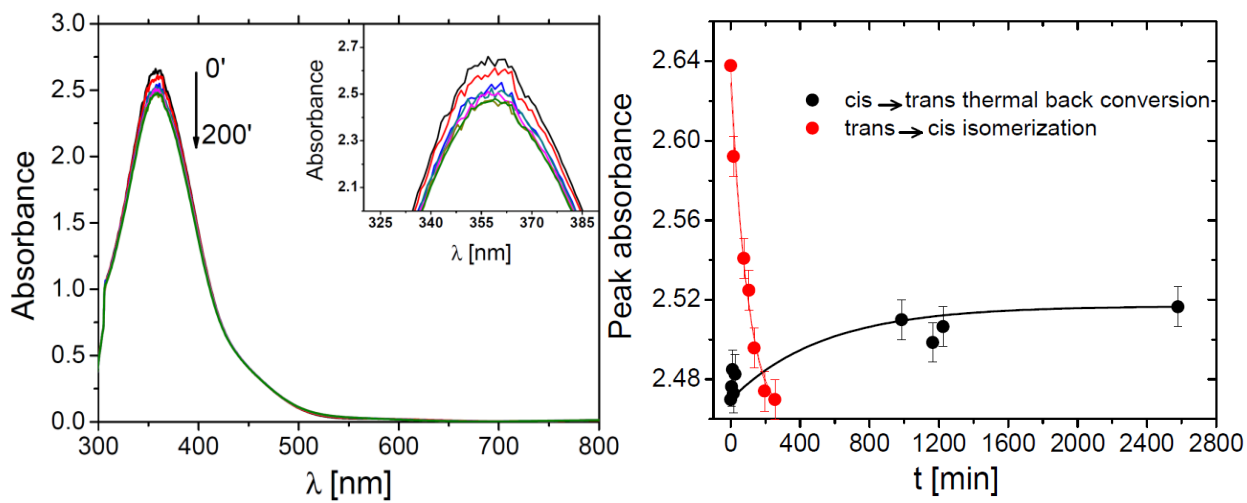

Figure S1: (a) absorption spectra during illumination. π-π* transition band centered at 358m and a magnification of the peak (inset) (b) Peak absorbance as a function of time for tran→cis isomerization during illumination and cis→trans thermal back conversion after turning off illumination. Continuous lines are plot only as guidelines.

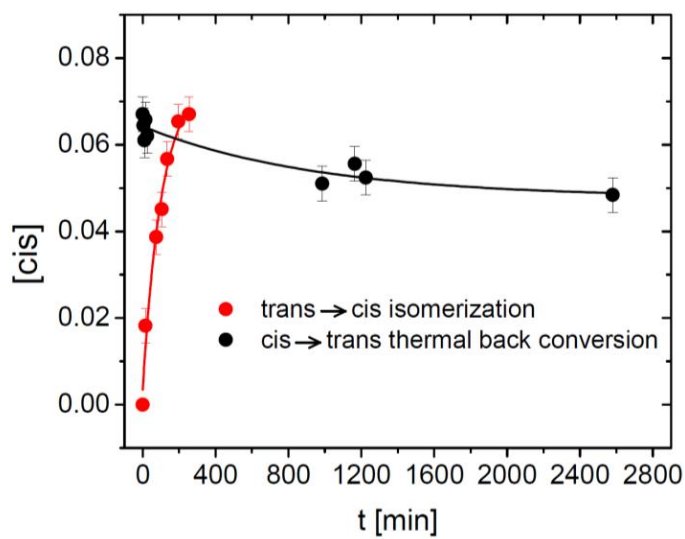

Figure S2: cis concentration [cis] as a function of time for trans→cis isomerization during illumination and cis→trans thermal back conversion after turning off illumination. Continuous lines are plot only as guidelines.

**Graphical abstract**

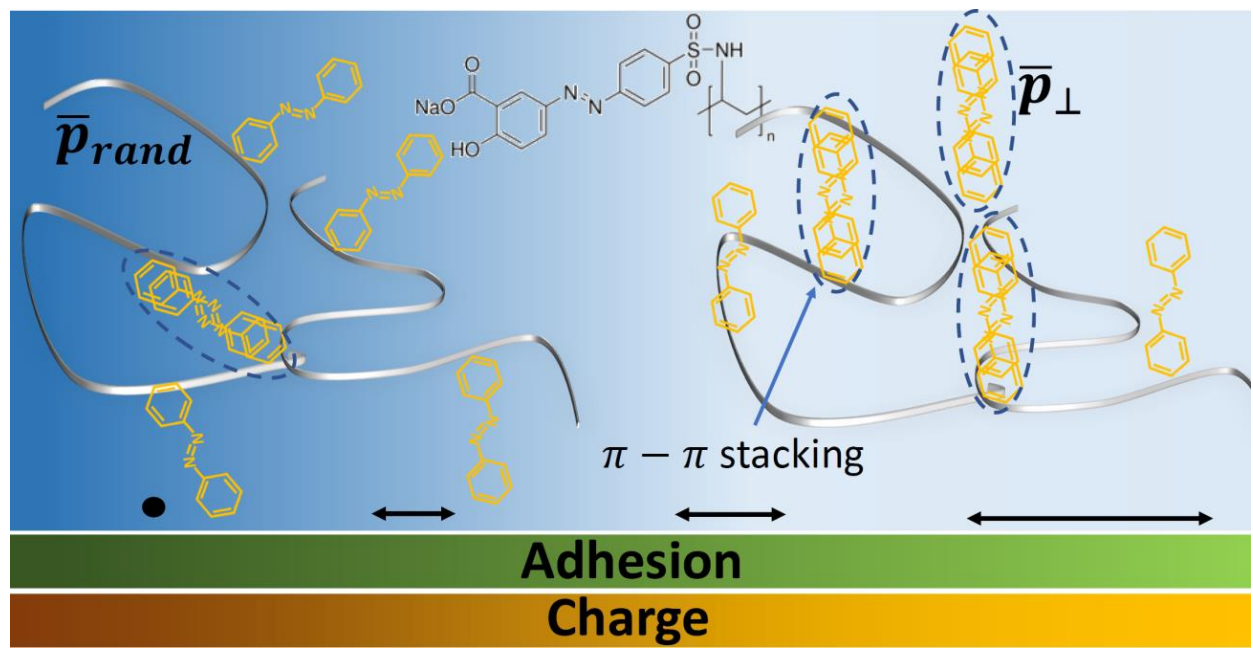

Adhesion and charge pattern formation on a SRG. The blue background is a section of the interference pattern used for illumination (In light blue the highest intensity). The azo compound orients with the transition dipole moment normal to the polarization vector (black arrows) in the brighter areas of the interference pattern, while remains random oriented in the darker areas. The azo compounds have higher probability to π stack in the bright areas where their planar structures are aligned (azo compounds in blue dashed ellipses). Higher adhesion and more electronegative charge distribution (represented with lighter colors in the green and yellow bars respectively), are found in the bright fringes of the interference pattern, where molecules have higher degree of organization, the π stacking of trans azobenzene is more probable, and the material polarization vector is bigger.

**Highlights**

- Supramolecular architectures with patterned surface properties are created in a parallel process.
- Adhesion and electrical charge patterns are generated during the SRG recording process.
- Azobenzene interactions enhance adhesion even in absence of SRG.
- π - stacking interaction contributes to generate a charge pattern.